\documentclass[conference]{IEEEtran}
\usepackage{amsm ath,amssymb,eucal,graphicx}
\usepackage{pst-plot}
\usepackage{epsfig}
\usepackage{amsthm}
\usepackage{amsfonts}
\usepackage{epsfig}
\usepackage{float}
\usepackage{pstricks}
\usepackage{multicol,subfigure}
\usepackage{mathcomSTEP,stmaryrd}

\theoremstyle{remark}
\newtheorem{rem}[thm]{Remark}
\theoremstyle{definition}

\begin{document}

\title{The Capacity of the Semi-Deterministic Cognitive Interference Channel with a Common Cognitive Message and
Approximate Capacity for the Gaussian Case
}

\author{
\authorblockN{Stefano Rini$^1$ and Carolin Huppert$^2$}
\authorblockA{$^1$ Institute for Communications Engineering, Technische Universit\"at M\"unchen, Germany, {\tt stefano.rini@tum.de}\\
$^2$ Institute of Communications Engineering, Ulm University, Germany, {\tt carolin.huppert@uni-ulm.de}}
}


\maketitle

\begin{abstract}

In this paper the study of the cognitive interference channel with a common message, a variation of the classical cognitive interference channel in which the cognitive message is decoded at both receivers.
We derive the capacity for the semi-deterministic channel in which the output at the cognitive decoder is a deterministic function of the channel inputs.
We also show capacity to within a constant gap and a constant factor for the Gaussian channel in which the outputs are linear combinations of the channel inputs plus an additive Gaussian noise term.
%
Most of these results are shown using an interesting transmission scheme in which the cognitive message, decoded at both receivers, is also
pre-coded against the interference experienced at the cognitive decoder.
The pre-coding of the cognitive message does not allow the primary decoder to reconstruct the interfering signal.
The cognitive message acts instead as a side information at the primary receiver when decoding its intended message.
\end{abstract}

{\IEEEkeywords
cognitive interference channel, superposition coding, binning, semi-deterministic channel, approximate capacity.
}

\section{Introduction}

Cognitive networks are transmission networks where the message of one user is known at multiple nodes.
The study of cognitive network was inspired by newfound abilities of smart radios  to overhear the transmission taking place over the channel and gather information about neighboring nodes \cite{goldsmith_survey}.
The information theoretical study of cognitive networks has so far focused on small networks with a limited number of users and messages.
A classical such model is the cognitive interference channel \cite{devroye_IEEE}: a channel where two sets of transmitter/receiver pair communicate over a shared channel, thus interfering with each other transmission. One of the encoders in the network has knowledge of only one of the messages to be transmitted
--the primary transmitter-- while the other node has knowledge of both messages --the cognitive transmitter.
The extra knowledge at the cognitive encoder models a smart and adaptable device that is able to acquire the primary message
from previous or simultaneous transmissions.
This model has been of great interest in the recent years, see \cite{RTDjournal1} and \cite{RTDjournal2} for a summary of the results,
and capacity is known for specific regimes.
Capacity is known in the ``weak  interference'' regime of \cite{WuDegradedMessageSet}, a regime where the interference created by the cognitive transmitter at the primary user is negligible and can be treated.
%
In the ``very strong interference'' regime of \cite{maric2005capacity} instead, capacity is achieved by having the primary receiver
 decode the interference created by the cognitive user and  strip it  from the received signal.
Capacity for the Gaussian case is also known in the  ``primary decodes cognitive'' regime of \cite{RTDjournal2},
where the cognitive message is decoded at both receivers and pre-coded against the interference created by the primary user at the cognitive decoder.
%

%
%
%

Despite of these and other results, available for the cognitive interference channel, capacity is not known in general.
In particular no capacity result is available for the channel where the cognitive output is a degraded version of the primary output.
The difficulty in determining capacity for this channel follows from the fact that pre-coding the cognitive message against the interference
from the primary user  has the effect of also canceling the primary signal at the primary receiver.
To gain new insights on this problem, we focus on a variation of the cognitive interference channel where the primary receiver also decodes the cognitive message.
%
%
%

\smallskip

\emph{ Paper organization and contributions}
\smallskip

{\bf $\bullet$ Sec. \ref{sec:Channel Model}: we introduce the model of the cognitive interference channel with a common cognitive message}, a variation of the cognitive interference channel where the primary decoder decodes both messages.

{\bf $\bullet$ Sec. \ref{sec:An Outer Bound for the Cognitive Interference Channel with a Common Cognitive Message}: we derive inner and outer bounds to the capacity region.} Both bounds are inspired by the cognitive interference channel in  the ``strong interference'' regime.  In this regime the primary decoder can reconstruct the channel output of the cognitive receiver after having decoded its intended message.

{\bf $\bullet$ Sec. \ref{sec:Capacity for the Semi-Deterministic Cognitive Interference Channel with a Common Cognitive Message}: we show the capacity for the semi-deterministic case}, that is the channel where the channel output at the cognitive decoder is a deterministic function of the channel inputs while the output at the primary receiver is any random function.


{\bf $\bullet$ Sec. \ref{sec:Capacity in the Very Strong Interference Regime}: we derive the capacity in the ``very strong interference regime'' and the
``primary decodes cognitive regime''.}
In the ``very strong interference'' regime there is no loss of optimality in having both decoders decode both messages while in the
``primary decodes cognitive'' regime there is no rate loss in having the primary receiver decodes the cognitive message.

{\bf $\bullet$ Sec. \ref{sec:Capacity to Within a Constant Gap and a Constant Factor}: we prove the capacity region of the Gaussian case to within a constant gap and a constant factor}.  That is we bound the difference between inner and outer bounds as well as the ratio between the two. These two results characterize the capacity region of the Gaussian case at large and small SNR respectively.


{\bf $\bullet$ Sec. \ref{sec:conclusion}: concludes the paper.}

\section{Channel Model}
\label{sec:Channel Model}

\begin{figure}
\centering
\includegraphics[width=8 cm]{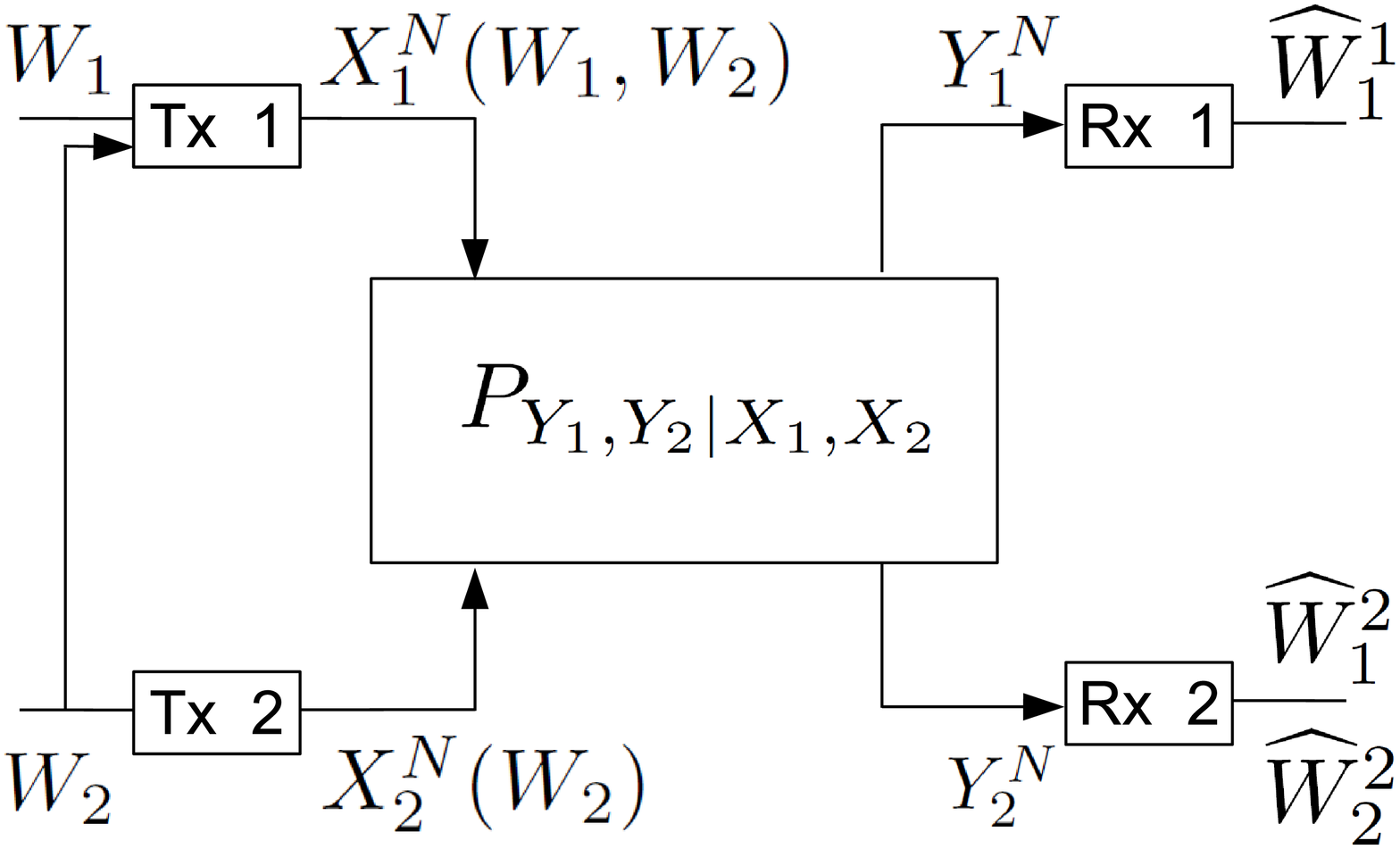}
\vspace{-1 cm}
\caption{The Cognitive InterFerence Channel with Common Cognitive Message (CIFC-CCM).}
\label{fig:channelModel}
\vspace{- .5 cm}
\end{figure}
The Cognitive InterFerence Channel with a Common Cognitive Message (CIFC-CCM), as shown in Fig. \ref{fig:channelModel}, is obtained from the classical Cognitive InterFerence Channel (CIFC)  by having the primary decoder decode both messages.
It consists of two transmitter-receiver pairs that exchange independent messages over a common channel. Transmitter $i$, $i\in\{1,2\}$, has discrete input alphabet $\Xcal_i$ and its receiver has discrete output alphabet $\Ycal_i$. The channel is assumed to be memoryless with transition probability $P_{Y_1,Y_2|X_1,X_2}$.
Encoder~2 wishes to communicate a message $W_2$ uniformly distributed on
$\Wcal_2 = [1: 2^{N R_2}]$ to decoder~2 in $N$ channel uses at rate $R_2$.
Similarly, encoder~1, wishes to communicate a message $W_1$ uniformly distributed on
$\Wcal_1 = [1: 2^{N R_1}]$ to both decoder~1 and decoder~2 in $N$ channel uses at rate $R_1$.
Encoder~1 (i.e., the cognitive user) knows its own message $W_1$ and that of encoder~2 (i.e. the primary user), $W_2$.
A rate pair $(R_1,R_2)$ is achievable if there exist sequences of encoding functions
\begin{align*}
X_1^N &= f_{X_1^N}(W_1, W_2), \;\;   f_{X_1^N} : \Wcal_1 \times \Wcal_2 \rightarrow {\cal X}_1^N, \\
X_2^N &= f_{X_2^N}(W_2), \;\; \;\;\;\;\;\;  f_{X_1^N} : \Wcal_2 \rightarrow {\cal X}_2^N,
\end{align*}
with corresponding sequences of decoding  functions
\begin{align*}
\widehat{W}_1^1 & =  f_{\widehat{W}_1^1}(Y_1^N), \;\;   f_{\widehat{W}_1^1} : {\cal Y}_1^N \rightarrow \Wcal_1, \\
\widehat{W}_1^2 & =  f_{\widehat{W}_1^2}(Y_1^N),   \;\; f_{\widehat{W}_1^2} : {\cal Y}_2^N \rightarrow \Wcal_1, \\
\widehat{W}_2^2 & =  f_{\widehat{W}_2^2}(Y_2^N),   \;\; f_{\widehat{W}_2^2} : {\cal Y}_2^N \rightarrow \Wcal_2.
\end{align*}
The capacity region is defined as the closure of the region of achievable $(R_1,R_2)$ pairs
 \cite{ThomasCoverBook}.
Standard strong-typicality is assumed; properties may be found in
 \cite{kramerBook}.



\medskip

\begin{figure}
\centering
\vspace{-.3 cm}
\includegraphics[width=8 cm]{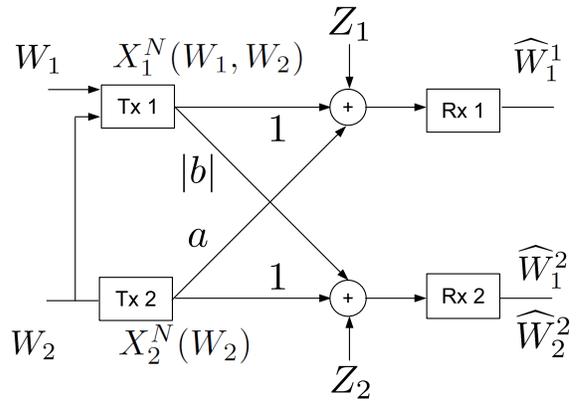}
\vspace{- .75 cm}
\caption{The Gaussian Cognitive Interference Channel with Common Cognitive Message.}
\label{fig:GaussianchannelModel}
\vspace{- .5 cm}
\end{figure}

In the following we focus in particular on Gaussian CIFC-CCM in Fig. \ref{fig:GaussianchannelModel}.
For this class of channels the input/output relationship
is:
\ea{
Y_1& =X_1 + \ a X_2 + Z_1, \label{eq:in/out gaussian CIFC-CCM} \\
Y_2& =X_2 + |b| X_1 + Z_2, \nonumber
}
for $a,b \in \Cbb$ and for $Z_i \sim \Ncal_{\Cbb}(0,1)$, where the $\Ncal_{\Cbb}$ indicates complex, circularly symmetric jointly Gaussian RV.
Moreover, the channel inputs are subject to the power constraints
\ea{
\Ebb \lsb X_i \rsb \leq P_i \ \ \ i \in \{1,2\}.
\label{eq:power constraint}
}
A channel where the outputs are obtained from a linear combination of the input plus an additional complex Gaussian term can be reduced to
the formulation in \eqref{eq:in/out gaussian CIFC-CCM} and \eqref{eq:power constraint} without loss of generality \cite[App. A]{RTDjournal2}.

\section{Outer and Inner Bounds for the Cognitive Interference Channel with a Common Cognitive Message}
\label{sec:An Outer Bound for the Cognitive Interference Channel with a Common Cognitive Message}

We start by deriving an outer bound for the capacity region of the general CIFC-CCM.
This outer bound is based on the results known for the cognitive interference channel in the ``strong interference'' regime.

\begin{thm}{\bf An Outer Bound for the CIFC-CCM}
\label{th:Outer Bound for the CIFC-CCM}

Any achievable region for the CIFC-CCM is contained in the region
\eas{
R_1         & \leq I(Y_1 ; X_1 | X_2 ),
\label{eq:Outer Bound for the CIFC-CCM R1} \\
R_1         & \leq I(Y_2 ; X_1 | X_2 ),
\label{eq:Outer Bound for the CIFC-CCM R1 weak} \\
R_1 + R_2   & \leq I(Y_2 ; X_1 , X_2),
\label{eq:Outer Bound for the CIFC-CCM sum rate 1}
}{\label{eq:Outer Bound for the CIFC-CCM}}
union over all the joint distributions of the channel inputs $P_{X_1,X_2}$.
\end{thm}

\begin{IEEEproof}
The outer bound in \eqref{eq:Outer Bound for the CIFC-CCM R1} was originally devised for the classical CIFC in \cite{WuDegradedMessageSet} and is valid for the
CIFC-CCM as well, since the cognitive decoder is decoding only the cognitive message.
The bound \eqref{eq:Outer Bound for the CIFC-CCM R1 weak}  is obtained from Fano's inequality  as
\pp{
& N R_1 - N \ep_{N}  \\
& \quad \quad  \leq I(Y_2^N ; W_1) \\
& \quad \quad  = I(Y_2^N ; W_1 | W_2) \\
& \quad \quad  = \sum_{ k=1}^N  H(Y_{2,k} | Y_{2, k+1}^N , W_2 , X_{2,k}) \\
& \quad \quad  \quad \quad  - H(Y_{2,k} | Y_{2, k+1}^N , W_1 , W_2, X_{1,k} , X_{2,k} ) \\
& \quad \quad  \leq \sum_{ k=1}^N  H(Y_{2,k} |  X_{2,k}) -  H(Y_{2,k} | X_{1,k} , X_{2,k} ) \\
& \quad \quad  \leq N I(Y_{2,Q}; X_{1,Q} |  X_{2,Q}, Q),
}
where $Q$ is the time sharing RV, uniformly distributed in the interval $\{1 \ldots N\}$.

The bound \eqref{eq:Outer Bound for the CIFC-CCM sum rate 1} is derived in a similar fashion:
\pp{
& N (R_1 + R_2 ) - N \ep_N  \\
& \quad \quad \leq I(Y_2^N ; W_1 , W_2)  \\
& \quad \quad  = \sum_{k=1}^N H(Y_{2, k} | Y_{2 , k+1}^N)\\
& \quad \quad  \quad \quad  - H(Y_{2, k} | Y_{2 , k+1}^N, W_1, W_2 , X_{1,k} , X_{2,k}) \\
& \quad \quad  \leq \sum_{k=1}^N I(Y_{2,k}; X_{1,k}, X_{2,k})\\
& \quad \quad  = N I(Y_{2,Q}; X_{1,Q}, X_{2,Q} |  Q).
}
All the bounds are decreasing in the time sharing RV $Q$ and thus it can be dropped.
\end{IEEEproof}

\begin{rem}
\label{rem:storng int outer bound}
The bound \eqref{eq:Outer Bound for the CIFC-CCM R1 weak} is redundant if
\ea{
I(X_1 ; Y_1 | X_2) \leq I(X_1 ; Y_2 |X_2),
\label{eq:strong interference}
}
for all the distributions $P_{X_1,X_2}$.
Condition \eqref{eq:strong interference} corresponds to the ``strong interference'' regime for the CIFC.
Thus, when dropping \eqref{eq:Outer Bound for the CIFC-CCM R1 weak} from the outer bound in Th.  \ref{th:Outer Bound for the CIFC-CCM}, one obtains the ``strong interference'' outer bound for the CIFC \cite{maric2005capacity}.
This outer bound  is capacity in the ``very strong interference'' regime for the general CIFC and is capacity in the ``primary decodes cognitive'' regime for the Gaussian CIFC.

\end{rem}

Rem. \ref{rem:storng int outer bound} formally defines the relationship between the CIFC in ``strong interference'' and the CIFC-CCM.
For a CIFC in the ``strong interference'' regime  there is no loss of optimality in having the primary decoder decodes both messages.
Under condition \eqref{eq:strong interference}, the rate of the cognitive message is not bounded by the decoding capabilities of the primary receiver.
%
For these reasons, the CIFC is equivalent to the CIFC-CCM when condition \eqref{eq:strong interference}  holds.
We avoid referring to condition \eqref{eq:strong interference}  as ``strong interference'' condition as one cannot properly define ``interference'' in the CIFC-CCM since the primary receiver is decoding both the cognitive message and the primary message.

The next theorem specializes the outer bound of Th. \ref{th:Outer Bound for the CIFC-CCM} to the Gaussian channel in \eqref{eq:in/out gaussian CIFC-CCM}.
\begin{cor}{\bf An Outer Bound for the Gaussian CIFC-CCM}
\label{cor:An Outer Bound for the Gaussian CIFC-CCM}

Any achievable region for the Gaussian CIFC-CCM is contained in the region
\eas{
R_1         & \leq  \Ccal(\al \min\{1 , |b|^2\}  P_1),
\label{eq:Outer Bound for the G-CIFC-CCM R1} \\
R_1 + R_2   & \leq \Ccal( P_2 + b^2 P_1 + 2 \sqrt{ \alb |b|^2 P_1P_2}),
\label{eq:Outer Bound for the G-CIFC-CCM sum rate 1}
}{\label{eq:Outer Bound for the G-CIFC-CCM}}
for $\Ccal(x)=\log(1+x)$.
\end{cor}
\begin{IEEEproof}
The outer bound is obtained from Th. \ref{th:Outer Bound for the CIFC-CCM} by noting that complex, circularly symmetric channel inputs maximize all the rate bounds simultaneously.
\end{IEEEproof}
We now develop an inner bound for the CIFC-CCM as follows: we rate-split the primary message in public and private part.
The primary private message is superposed to the public primary message; the cognitive message is superposed to the primary public message and
pre-coded against the primary private message.

\begin{thm}{\bf Inner Bounds for the CIFC-CCM}
\label{th:Inner Bounds for the CIFC-CCM}
The following region is achievable for a general CIFC-CCM
\eas{
R_1 & \leq I(Y_1 ; U_{1c} | U_{2c})-I(U_{1c}; X_2 | U_{2c}), \\
R_1 & \leq
I(Y_2; X_1 | X_2, U_{2c}), \\
R_1 + R_2 & \leq I(Y_1; U_{1c}, U_{2c})+I(Y_2; X_2 | U_{1c}, U_{2c}),\\
R_1+R_2 & \leq I(Y_2 ; X_1 , X_2), \\
2 R_1 + R_2 & \leq
I(Y_1 ; U_{1c} , U_{2c}) +  I(Y_2; X_1 , X_2 | U_{2c})\nonumber \\
&\quad \quad -I(U_{1c}; X_2 | U_{2c}),
\label{eq:inner bound 2R1+R2}
}{\label{eq:inner bound}}
for any distribution that factors as
$P_{U_{1c},U_{2c},X_1,X_2}$.
\end{thm}

\begin{IEEEproof}
The common cognitive message is embedded in the codeword $U_{1c}^N$, with rate $R_{1c}$, while the primary common message in the codeword $U_{2c}^N$, with rate $R_{2c}$ and the primary private message in the codeword $X_2^N$ with rate $R_{2p}$.
The codeword $U_{1c}^N$ is binned against $X_2^N$ and the codewords $U_{1c}^N$ and  $X_2^N$ are both superposed to $U_{2c}^N$.
The channel input $X_1^N$ is finally obtained as a deterministic function of $U_{1c}^N,U_{2c}^N$ and $X_2^N$.
From \cite{rini2011achievable} we obtain the achievable region
\eas{
\Ro_{1c} \quad \quad \ \ & \geq I(U_{1c} ; X_2 | U_{2c}) \\
R_{1c}+\Ro_{1c} + R_{2c}   & \leq I(Y_1; U_{1c} , U_ {2c})\\
R_{1c}+\Ro_{1c}        \quad \quad \ \    & \leq I(Y_1; U_{1c} |U_{2c}) \\
R_{2c}+ R_{1c}+\Ro_{1c} + R_{2p} & \leq I(Y_2 ; U_{1c}, U_{2c},X_2) \nonumber\\
& \quad \quad  + I(U_{1c}; X_2 |U_{2c})\\
R_{1c}+\Ro_{1c} + R_{2p} & \leq I(Y_2 ; U_{1c}, X_2| U_{2c})  \nonumber \\
& \quad \quad + I(U_{1c}; X_2 |U_{2c})\\
R_{1c}+\Ro_{1c} \quad \quad \ \   & \leq I(Y_2 ; U_{1c} | X_2, U_{2c})  \nonumber \\
& \quad \quad  + I(U_{1c}; X_2 |U_{2c}) \\
R_{2p} & \leq I(Y_2 ; X_2| U_{1c}, U_{2c})  \nonumber \\
& \quad \quad   + I(U_{1c} ; X_2 | U_{2c}).
}{\label{eq:general inner bound before FME}}
By applying the FME with
\ea{
R_1  = R_{1c}, \quad \quad R_2 = R_{2c} + R_{2p},
}
we obtain the region  in  \eqref{eq:inner bound}.
\end{IEEEproof}

\begin{figure}
\centering
\includegraphics[width=6 cm]{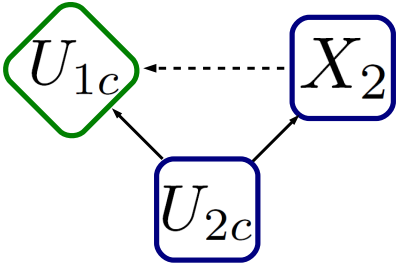}
\vspace{-.5   cm}
\caption{The chain graph representation of the achievable scheme in Th. \ref{th:Inner Bounds for the CIFC-CCM}.}
\label{fig:CGRAS}
\vspace{-.6 cm}
\end{figure}

The chain graph representation of the achievable scheme \cite{rini2011achievable} in Th. \ref{th:Inner Bounds for the CIFC-CCM} is provided in Fig \ref{fig:CGRAS}:
the blue boxes represent codewords associated with the primary message while the green diamond represents the cognitive message. Solid lines represent superposition coding and dotted lines binning.

Note that the achievable region in \cite[Th. 7]{RTDjournal1} for the CIFC has less bound that the region in Th. \ref{th:Inner Bounds for the CIFC-CCM}.
This follows from the fact that the correct decoding of the cognitive message is not required in the CIFC and some error events in the scheme of Th. \ref{th:Inner Bounds for the CIFC-CCM} are not errors in the achievable scheme of \cite[Th. 7]{RTDjournal1}.

\section{Capacity for the Semi-Deterministic Cognitive Interference Channel with a Common Cognitive Message}
\label{sec:Capacity for the Semi-Deterministic Cognitive Interference Channel with a Common Cognitive Message}

The semi-deterministic CIFC-CCM is a general CIFC-CCM where the channel output at the cognitive decoder is a deterministic function of the channel inputs, i.e.
\ea{
Y_1 = f_{Y_1} (X_1,X_2)
\label{eq:det condition Y_1}
}
while the primary output is any random function of the inputs.
When condition \eqref{eq:det condition Y_1} holds, binning at the cognitive transmitter can fully pre-cancel the effect of the interference at the cognitive receiver thus making \eqref{eq:Outer Bound for the CIFC-CCM R1} achievable.


\begin{thm}{\bf Capacity of the Semi-Deterministic CIFC-CCM}
\label{th:Capacity of the semi deterministic CIFC-CCM}
The capacity of the semi-deterministic channel is
\eas{
R_1 & \leq  H(Y_1|X_2) \\
R_1 & \leq  I(Y_2 ; X_1 |  X_2) \\
R_1 + R_2 & \leq I(Y_2; X_1, X_2)
}{\label{eq:Capacity of the semi deterministic CIFC-CCM}}
union over all the distributions $P_{X_1,X_2}$.
\end{thm}

\begin{IEEEproof}
Consider the transmission scheme in Th. \ref{th:Inner Bounds for the CIFC-CCM} for $U_{2c}=\emptyset$ to obtain the region
\eas{
R_1 & \leq I(Y_1 ; U_{1c}) - I(U_{1c}; X_2)
\label{eq:scheme E R1 bound }\\
R_1 & \leq I(Y_2 ; X_1| X_2)
\label{eq:scheme E R1 bound 2}\\
R_1 + R_2 & \leq I(Y_2 ; X_1, X_2)
\label{eq:scheme E sum rate Y1}\\
R_1 + R_2 & \leq I(Y_1; U_{1c}) + I(Y_2 ; X_2 | U_{1c})
\label{eq:scheme E extra sum rate Y1}
}{\label{eq:scheme E}}
where we have dropped \eqref{eq:inner bound 2R1+R2} since, with $U_{2c}=\emptyset$,
\ea{
& {\rm RHS-}{\eqref{eq:scheme E R1 bound }}+ {\rm RHS-}{\eqref{eq:scheme E sum rate Y1}} ={\rm RHS-}\eqref{eq:inner bound 2R1+R2}.
}
For the assignment $U_{1c}=Y_1$, which is possible given \eqref{eq:det condition Y_1}, the inner bound in \eqref{eq:scheme E} coincides with
\eqref{eq:Capacity of the semi deterministic CIFC-CCM} since
\eas{
& {\rm RHS-}{\eqref{eq:scheme E extra sum rate Y1}}   = H(Y_1)+ H(Y_2 | Y_1)  +H(Y_2 | X_1, X_2)  \nonumber \\
& = I(Y_2 ; X_1, X_2) + H(Y_1 |Y_2)  \geq {\rm RHS-}{\eqref{eq:scheme E sum rate Y1}}.
}
which is also equivalent to the outer bound.
\end{IEEEproof}

%
%

\section{Capacity in the ``Very Strong Interference'' Regime and the ``Primary Decodes Cognitive'' Regime}
\label{sec:Capacity in the Very Strong Interference Regime}
%
%
Capacity in the ``very strong interference'' regime  for the CIFC is achieved by having both decoders decode both messages and by superposing the cognitive message over the primary message \cite{maric2005capacity}.
This strategy achieves capacity also for a class of CIFC-CCM that we also term  ``very strong interference'' regime.
This definition is not fully accurate since the primary receiver decodes both messages, but is coherent with the CIFC literature.
%
%
%
\begin{thm}{\bf Capacity in the Very Strong Interference Regime}
\label{th:Capacity in the Very Strong Interference Regime}
If
\ea{
I(Y_2 ; X_1 , X_2) \leq  I(Y_1 ; X_1 , X_2),
\label{eq:very strong interference}
}
the region in \eqref{eq:Outer Bound for the CIFC-CCM} is capacity.
\end{thm}
\begin{IEEEproof}
%
Consider the scheme in \eqref{eq:inner bound}. For $X_2=U_{2c}$ and $U_{1c}=X_1$ the achievable region is
\eas{
R_1 & \leq  I(Y_1 ; X_1 | X_2) \\
R_1 & \leq I(Y_2 ; X_1 | X_2) \\
R_1 + R_2 & \leq I(Y_1 ; X_1 , X_2)
\label{eq:scheme D Y1 sum rate}\\
R_1 + R_2 & \leq I(Y_2; X_1 , X_2).
}
Under condition \eqref{eq:very strong interference} the bound in \eqref{eq:scheme D Y1 sum rate} can be eliminated from the inner bound and the inner bound  is then equivalent to the outer bound in \eqref{eq:Outer Bound for the CIFC-CCM}.
\end{IEEEproof}

\begin{rem}
The ``very strong interference'' regime for the CIFC is defined by condition \eqref{eq:very strong interference} and \eqref{eq:strong interference}.
However, condition \eqref{eq:strong interference} is not required to prove capacity for the CIFC-CCM.
\end{rem}
%
%

The following corollary states the result of Th. \ref{th:Capacity in the Very Strong Interference Regime} for the Gaussian case in \eqref{eq:in/out gaussian CIFC-CCM}.

\begin{cor}{\bf Capacity for the Gaussian CIFC-CCM in the Very Strong Interference Regime}
\label{cor:Capacity for the Gaussian CIFC-CCM in the very strong interference regime}
If
\ea{
(|a|^2-1)P_2 -(|b|^2-1)P_1 - 2 |a - |b||\sqrt{P_1 P_2} \geq 0
\label{eq:very strong interference gaussian}
}
the capacity of the Gaussian CIFC-CCM is given  by \eqref{eq:Outer Bound for the G-CIFC-CCM}.
\end{cor}
\begin{IEEEproof}
%
Condition \eqref{eq:very strong interference gaussian} is derived from \eqref{eq:very strong interference} for the Gaussian model in \eqref{eq:in/out gaussian CIFC-CCM}. Details can be found in \cite[App. B]{RTDjournal2}.
\end{IEEEproof}

We can extend the ``primary decodes cognitive'' regime of \cite{RTDjournal2} to the Gaussian CIFC-CCM:

\begin{thm}{\bf The Primary Decodes Cognitive Interference Regime for the CIFC-CCM}
\label{th:Primary Decodes Cognitive Interference Regime for the CIFC-CCM}
If
\ea{
P_2 \labs 1- a |b|\rabs^2 & \geq (|b|^2-1)(1 + P_1 + |a|^2 P_2 ) - P_1 P_2 \labs 1 - a |b|\rabs^2
\nonumber \\
P_2 \labs 1- a |b|\rabs^2 & \geq (|b|^2-1)(1 + P_1 + |a|^2 P_2 +2 \Re\{a\}\sqrt{P_1 P_2 }),
\label{eq:PDC}
}
then \eqref{eq:Outer Bound for the G-CIFC-CCM} is the  capacity of the Gaussian CIFC-CCM.
\end{thm}
\begin{IEEEproof}
Consider the scheme in \eqref{eq:scheme E} with the assignment
\eas{
X_i \sim \Ncal(0, P_i) \quad i \in \{1,2\} \\
U_{1c}= X_1 + \f{\al P}{\al P_1+1} a X_2,
}
which yields the achievable region
\eas{
R_1 & \leq \Ccal  \lb \al \min\{1,|b|^2\}P_1\rb
\label{eq:schemeE gaussian R1} \\
R_1 + R_2 & \leq \Ccal\lb |b|^2 P_1 + P_2  + 2 \sqrt{\alb |b|^2 P_1 P_2}\rb  \nonumber \\
 & \quad \quad  + f \lb a + \sqrt{\f{\al P_1}{P_2}};1 ; \f{\al P_1}{\al P_1 +1} \rb  \nonumber  \\
  & \quad \quad - f \lb \f 1 {|b|^2} + \sqrt{\f{\al P_1}{P_2}}, \f 1 {|b|^2} ; \f{\al P_1}{\al P_1 +1} \rb
\label{eq:schemeE gaussian R2} \\
R_1+R_2 & \leq \Ccal \lb |b|^2 P_1 + P_2  +2 \sqrt{\alb |b|^2 P_1 P_2 }\rb,
\label{eq:schemeE gaussian R1+R2}
}{\label{eq:schemeE gaussian}}
for
$$
f (h, \sgs, \lambda)= \log \lb \f{\sgs+ \al P_1 }{ \sgs + \f{\al P_1 |h|^2 P_2}{\al P_1 |h|^2P_2+ \sgs} \labs \f{\la}{\la_{\rm Costa}(h,\sgs)}-1\rabs } \rb
$$
and
$$
\la_{\rm Costa} (h, \sgs)= \f{\al P_1}{\al P_1 +\sgs}h.
$$
%
This scheme achieves capacity when \eqref{eq:schemeE gaussian R2} is larger than \eqref{eq:schemeE gaussian R1+R2}.
The conditions were determined in \cite{RTDjournal2} to prove the ``primary decodes cognitive'' regime for the CIFC.
\end{IEEEproof}
\begin{rem}
The ``primary decodes cognitive regime'' for the CIFC is defined by condition \eqref{eq:PDC} and condition \eqref{eq:strong interference} which is given by $|b| \geq 1$ in the Gaussian case.
Condition \eqref{eq:strong interference} is not required to prove capacity for the CIFC-CCM.
The capacity of the Gaussian CIFC-CCM for $|b| \leq 1$ is given by Corollary \ref{th:Primary Decodes Cognitive Interference Regime for the CIFC-CCM}.
\end{rem}
\section{Capacity to Within a Constant Gap and a Constant Factor}
\label{sec:Capacity to Within a Constant Gap and a Constant Factor}

\begin{thm}{\bf Capacity to within 1.87 bits}
For any Gaussian CIFC-CCM, the outer bound region in \eqref{eq:Outer Bound for the G-CIFC-CCM} can be achieved to within 1.87 bits/s/Hz.
\end{thm}

\begin{IEEEproof}
Capacity is known for $|b| \leq 1$. The achievability of the outer bound in \eqref{eq:Outer Bound for the G-CIFC-CCM} to within  $1.87$ bits/s/Hz for $|b|>1$  using the scheme in \eqref{eq:scheme E} is shown in \cite{rini2010capacity}.
\end{IEEEproof}

\begin{thm}{\bf Capacity to within a factor 2}
For any Gaussian CIFC-CCM, the outer bound region in \eqref{eq:Outer Bound for the G-CIFC-CCM} can be achieved to within a factor 2.
\end{thm}

\begin{IEEEproof}
Capacity is known for $|b| \leq 1$.  The achievability of the outer bound in \eqref{eq:Outer Bound for the G-CIFC-CCM} to within  a factor $2$ for $|b|>1$  by a simple time division scheme is shown in \cite{RTDjournal2}.
%
\end{IEEEproof}

\begin{figure}
\centering
\includegraphics[width=11 cm]{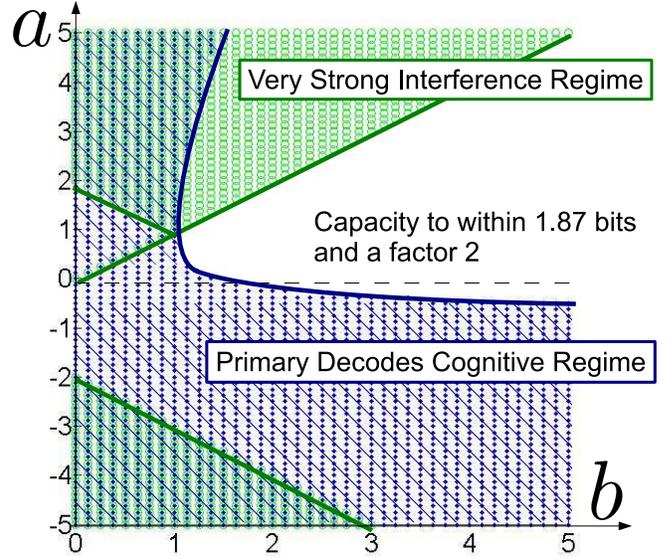}
\vspace{-1 cm}
\caption{Capacity for Gaussian CIFC-CCM.}
\label{fig:Gaussian Plot}
\vspace{-.25 cm}
\end{figure}

A plot of the capacity results available for the Gaussian CIFC-CCM is depicted in Fig. \ref{fig:Gaussian Plot}: in the $a \times b$ plane we plot the ``strong interference'' regime (green, right-hatched area) and the ``primary decodes cognitive regime'' (blue, left-hatched area).

\section{Conclusion}
\label{sec:conclusion}

 In this paper we study a variation of the classical cognitive interference channel where the primary receiver decodes both messages.
 This channel is related to the cognitive interference channel in the ``strong interference'' regime and many results for this channel apply to the model under consideration.
 We derive the capacity for the semi-deterministic case, where the cognitive output is a deterministic function of the channel inputs.
 We also show capacity in the ``strong interference'' regime, where there is no rate loss  in having both receivers decode both messages.
 For the Gaussian channel, we determine capacity in the ``primary decodes cognitive'' regime and determine capacity to within a constant gap and to within a constant factor.

\bibliographystyle{plain}
\bibliography{steBib1}

\begin{thebibliography}{10}

\bibitem{ThomasCoverBook}
T.~Cover and J.~Thomas.
\newblock {\em Elements of Information Theory}.
\newblock Wiley-Interscience, New York, 1991.

\bibitem{devroye_IEEE}
N.~Devroye, P.~Mitran, and V.~Tarokh.
\newblock Achievable rates in cognitive radio channels.
\newblock {\em IEEE Trans. Inf. Theory}, 52(5):1813--1827, May 2006.

\bibitem{goldsmith_survey}
A.~Goldsmith, S.A. Jafar, I.~Maric, and S.~Srinivasa.
\newblock Breaking spectrum gridlock with cognitive radios: An information
  theoretic perspective.
\newblock {\em Proc.of the IEEE}, 2009.

\bibitem{kramerBook}
G.~Kramer.
\newblock {\em Topics in Multi-User Information Theory}.
\newblock Foundations and Trends in Communications and Information Theory. Vol.
  4: No 4�5, pp 265-444, 2008.

\bibitem{maric2005capacity}
I.~Maric, R.~Yates, and G.~Kramer.
\newblock {The capacity region of the strong interference channel with common
  information}.
\newblock In {\em Proc. Asilomar Conferenece on Signal, Systems and Computers},
  pages 1737--1741, Nov. 2005.

\bibitem{rini2011achievable}
S.~Rini.
\newblock An achievable region for a general multi-terminal network and the
  corresponding chain graph representation.
\newblock {\em Arxiv preprint arXiv:1112.1497}, 2011.

\bibitem{rini2010capacity}
S.~Rini, D.~Tuninetti, and N.~Devroye.
\newblock {The capacity region of {G}aussian cognitive radio channels to within
  1.87 bits}.
\newblock In {\em Proc. IEEE Inf. Theory Workshop}, January 2010.

\bibitem{RTDjournal1}
S.~Rini, D.~Tuninetti, and N.~Devroye.
\newblock {New Inner and Outer Bounds for the Memoryless Cognitive Interference
  Channel and some new Capacity Results}.
\newblock {\em IEEE Trans. Inf. Theory}, 57(7):4087 -- 4109, July 2011.

\bibitem{RTDjournal2}
S.~Rini, D.~Tuninetti, and N.~Devroye.
\newblock Inner and outer bounds for the {G}aussian cognitive interference
  channel and new capacity results.
\newblock {\em IEEE Trans. Inf. Theory}, 2012.
\newblock accepted for publication.

\bibitem{WuDegradedMessageSet}
W.~Wu, S.~Vishwanath, and A.~Arapostathis.
\newblock Capacity of a class of cognitive radio channels: Interference
  channels with degraded message sets.
\newblock {\em IEEE Trans. Inf. Theory}, 53(11):4391--4399, Nov. 2007.

\end{thebibliography}

\end{document}